\documentclass[copyright,creativecommons]{eptcs}
\providecommand{\event}{PLACES 2016} 
\usepackage{breakurl}             

\usepackage[nounderscore]{syntax}
\usepackage{amsmath}
\usepackage{amssymb}
\usepackage{array}
\usepackage{xypic}
\usepackage{tabularx}
\usepackage{pdfpages}

\newcommand{\ignore}[1]{ }

\renewcommand{\b}[1]{\textbf{#1}}
\newcommand{\m}[1]{\mathit{#1}}

\newcommand{\val}[1]{C_{#1}[x].\mathit{val}}
\newcommand{\state}[1]{C_{#1}[x].\mathit{state}}
\newcommand{\clean}{\texttt{clean}}
\newcommand{\dirty}{\texttt{dirty}}
\newcommand{\dom}[1]{\mathit{dom}(C_{#1})}

\newcommand{\cacheup}[2]{\xrightarrow[#1 \to #2]{x}}
\newcommand{\storeup}[1][i]{\xrightarrow[#1 \to S]{x}}
\newcommand{\eviction}[1][i]{\xrightarrow[#1\upharpoonright]{x}}
\newcommand{\reconf}[2]{{}_{#1}\looparrowright^{#2}}

\newtheorem{prop}{Property}
\newtheorem{proposition}{Proposition}
\newtheorem{definition}{Definition}
\newtheorem{theorem}{Theorem}

\makeatletter
\def\overbracket#1{\mathop{\vbox{\ialign{##\crcr\noalign{\kern3\p@}
      \downbracketfill\crcr\noalign{\kern3\p@\nointerlineskip}
      $\hfil\displaystyle{#1}\hfil$\crcr}}}\limits}
\def\underbracket#1{\mathop{\vtop{\ialign{##\crcr
      $\hfil\displaystyle{#1}\hfil$\crcr\noalign{\kern3\p@\nointerlineskip}
      \upbracketfill\crcr\noalign{\kern3\p@}}}}\limits}
\def\overparenthesis#1{\mathop{\vbox{\ialign{##\crcr\noalign{\kern3\p@}
      \downparenthfill\crcr\noalign{\kern3\p@\nointerlineskip}
      $\hfil\displaystyle{#1}\hfil$\crcr}}}\limits}
\def\underparenthesis#1{\mathop{\vtop{\ialign{##\crcr
      $\hfil\displaystyle{#1}\hfil$\crcr\noalign{\kern3\p@\nointerlineskip}
      \upparenthfill\crcr\noalign{\kern3\p@}}}}\limits}
\def\downparenthfill{$\m@th\braceld\leaders\vrule\hfill\bracerd$}
\def\upparenthfill{$\m@th\bracelu\leaders\vrule\hfill\braceru$}
\def\upbracketfill{$\m@th\makesm@sh{\llap{\vrule\@height3\p@\@width.7\p@}}%
  \leaders\vrule\@height.7\p@\hfill
  \makesm@sh{\rlap{\vrule\@height3\p@\@width.7\p@}}$}
\def\downbracketfill{$\m@th
  \makesm@sh{\llap{\vrule\@height.7\p@\@depth2.3\p@\@width.7\p@}}%
  \leaders\vrule\@height.7\p@\hfill
  \makesm@sh{\rlap{\vrule\@height.7\p@\@depth2.3\p@\@width.7\p@}}$}
\makeatother
\newcommand{\reduct}[1]{\raise3.5pt\hbox{\scriptsize$\ulcorner$}\kern-3.0pt#1\kern-2.6pt\raise3.5pt\hbox{\scriptsize$\urcorner$}}

\makeatletter
\newcommand{\rd}[1][x]{%
  \@ifnextchar[{\rd@ii[#1]}{\rd@ii[#1][v]}
}
\def\rd@ii[#1][#2]{%
  \@ifnextchar[{\rd@iii[#1][#2]}{\rd@iii[#1][#2][]}
}
\def\rd@iii[#1][#2][#3]{%
  {\texttt{rd{#3}}_{#1}^{#2}}
}
\makeatother

\makeatletter
\renewcommand{\wr}[1][x]{%
  \@ifnextchar[{\wr@ii[#1]}{\wr@ii[#1][v]}
}
\def\wr@ii[#1][#2]{%
  {\texttt{wr}_{#1}^{#2}}
}
\makeatother

\makeatletter
\newcommand{\srd}[1][x]{%
  \@ifnextchar[{\srd@ii[#1]}{\srd@ii[#1][v]}%
}
\def\srd@ii[#1][#2]{%
  {\texttt{srd}_{#1}^{#2}}
}
\makeatother

\makeatletter
\newcommand{\swr}[1][x]{%
  \@ifnextchar[{\swr@ii[#1]}{\swr@ii[#1][v]}
}
\def\swr@ii[#1][#2]{%
  {\texttt{swr}_{#1}^{#2}}
}
\makeatother

\newcommand{\trans}[1]{\xrightarrow{(a_{#1}, i_{#1})}}
\newcommand{\ptrans}[2][i]{\xrightarrow{(#2,#1)}}
\newcommand{\strans}{\xrightarrow{*}}
\newcommand{\seqtrans}[1]{\stackrel{#1}{\Longrightarrow}}

\newcommand{\N}{\ensuremath{\mathbb{N}}}
\newcommand{\C}{\ensuremath{\mathbb{C}}}

\newcommand{\assign}{\ensuremath{\mathrel{\mathop:}=}}

\title{Program Execution on Reconfigurable Multicore Architectures}
\author{Sanjiva Prasad
\institute{Indian Institute of Technology Delhi\\ New Delhi, India}
\email{sanjiva@cse.iitd.ac.in}
}
\def\titlerunning{Reconfigurable Multicore Architectures}
\def\authorrunning{S. Prasad}
\begin{document}
\maketitle

\begin{abstract}
Based on the two observations that diverse applications perform better on different multicore  architectures, and that different phases of an application may have vastly different resource requirements,  Pal {\em et al.}  proposed a novel reconfigurable hardware approach for executing multithreaded programs. 
Instead of mapping a concurrent program to a fixed architecture,  the architecture adaptively reconfigures itself to meet the application's concurrency and communication requirements, yielding significant improvements in performance.  
Based on our earlier abstract operational framework for multicore execution with hierarchical memory structures, we describe execution of multithreaded programs on reconfigurable architectures that  support a variety of clustered configurations. 
Such reconfiguration may not preserve the semantics of programs due to the possible introduction of race conditions arising from concurrent accesses to shared memory by threads running on the different cores.  
We present an intuitive partial ordering notion on the cluster configurations, and show that the semantics of multithreaded programs is always preserved for reconfigurations ``upward'' in that ordering, whereas semantics preservation for arbitrary reconfigurations can be guaranteed for well-synchronised programs.  
We further show that a simple approximate notion of efficiency of execution on the different configurations can be obtained using the notion of amortised bisimulations, and extend it to dynamic reconfiguration. 
\end{abstract}

\section{Introduction}\label{INTRODUCTION}

The traditional approach to multiprocessing is to map multithreaded applications or multiprogram workloads onto a chosen multicore architecture.
Pal {\it et al.}  showed that due to the diversity of software applications, this ``one architecture fits all'' approach often yields sub-optimal performance \cite{PalPP-ISPA12}.   
For instance, programs with mostly independent threads having little communication perform well on symmetric multiprocessors (SMP) whereas those with more communication and shared variables perform better on chip multiprocessors (CMP).
Indeed, even a single application can exhibit vastly diverse resource requirements during different phases of its execution. 
 Accordingly, those authors identified different reconfiguration parameters (e.g., number of cores, cache size and cache sharing) and proposed a reconfigurable multicore tile-based architecture which supports dynamic {\em adaptability of the multicore hardware to the software's resource requirements} \cite{PalPP-JPDC14}, obtaining significant performance improvements.  
 The {\em hardware morphs itself} to a configuration that delivers better performance for that particular phase of program execution (using heuristics to detect such phase changes).   
The overhead for reconfiguration is usually significantly lower than the performance benefits. 
However, it is not entirely obvious whether such a dynamic reconfiguration preserves the intended semantics of the application with respect to a reference architecture, nor is there a theoretical framework for comparing the performance benefits. 
We believe there should be a formal basis for dynamically reconfigurable multiprocessors,  which constitute an innovative technology trend. 

In this paper, we present an operational account of execution of multithreaded programs  on {\em dynamically reconfigurable} multicore architectures, which to our knowledge is new.  
Examining how a variety of {\em cluster-based} architectural configurations partition cores and share cache, we see that partition refinements provide a natural partial ordering on cluster configurations.  
We also find that instead of requiring different operational formulations for each configuration, our earlier work \cite{JoshiP-caches} provides a {\em uniform} abstract operational semantics, whence we can both compare execution behaviour semantics as well as express dynamic reconfigurability. 
Leveraging results about abstract cache models from our work \cite{JoshiP-caches}, which followed the seminal approach of Boudol and Petri \cite{BoudolP-relaxed} of showing that ``well-synchronised programs have the same semantics in relaxed memory models as on sequentially consistent models'', we show that dynamic reconfiguration is semantics-preserving for such data-race-free programs. 
Further, by associating an approximate cost with each operation, we adapt Kiehn and Arun-Kumar's notion of {\em amortised bisimulation} \cite{KiehnA-amortized} to obtain a framework for comparing performance on the various architectural configurations and reasoning about the benefits of dynamic reconfiguration. 

It should be clarified at the outset that we are presenting execution semantics at the {\em architectural} level,  {\em below the program or OS level} where threads are assigned to cores.   
Reconfiguration happens dynamically during program execution, and is {\em outside program control}. 
In this respect,  this work differs from that of Krishnan \cite{Krishnan-multi,Krishnan-archCCS} and also the large body of work related to assigning threads to cores (which has anyway become a less pressing issue in multicore systems).
For simplicity, we confine our study to homogeneous core architectures, and the reconfiguration parameters to core clustering and cache fusion/splitting, and do not consider other parameters such as core fusion/splitting, core allocation, management of power and clocking, and cache allocation.  

\section{The Reference Model}\label{SPEC}
The reference architectural model with respect to which we compare the semantics of program execution consists of a collection of cores connected via a bus to a single shared memory module.   
Under the assumption of having the requisite number of cores,  this reference model will exhibit behaviours consistent with {\em pomset} semantics \cite{Pratt84} of multiple sequential processes with a {\em sequentially consistent} shared memory \cite{Lamport79}.    In a sequentially consistent memory model, writes and reads are atomic operations, and occur in program order within a thread.

Execution states are written as $\C = (S,P)$, where $S$ is the shared {\em store} and $P$ is a vector of threads. For simplicity, we assume each thread runs on its own core, with   $P_i$ denoting the $i^{th}$ thread.  
The operational semantics are given in Figure \ref{specificationsem}.  
Since our focus is on the observable actions on shared memory at the architectural level,  only the transitions relating to reading and writing from memory are shown, eliding transitions for instructions not involving the store.   
Since the bus enforces {\em mutually exclusive access} to the  memory module, we adopt an interleaving  view of execution.    
	An obvious alternative approach would have been to consider {\em synchronous transitions}, labelled by {\em vectors of actions}, the components of which are contributed by each core.  
	However, since certain cores may idle (e.g., for power efficiency reasons) and since we are not making any assumptions about  clock synchronicity, that  approach would not be appropriate.    

\begin{figure}[t]
	\centering
	\begin{tabularx}{\textwidth}{>{\(}l<{\)}}
		({read})~~~~(S, P_i[x]) \ptrans{\rd}_S (S, P_i[v]) ~~~~~ \textit{where } S(x) = v \\ 
		({write})~~~(S, P_i[x $\assign$ v]) \ptrans{\wr}_S (S[x \gets v], P_i[()])   \\
	\end{tabularx}
	\caption{Specification semantics for read and write in the reference architecture}\label{specificationsem}
\end{figure}
Transitions related to accessing the store are of the form $\xrightarrow{(a,i)}$, where $i$ is used to indicate that the transition is for thread $P_i$ (or core $i$), and $a$ denotes the action. 
The possible actions are: $\rd$ (the value $v$ is read from variable $x$) and $\wr$ (the value $v$ is written to the variable $x$), apart from the reductions (labelled $\tau$) not involving the store.  
When we do not care what value was read/written, we use $\rd[x][]$ and $\wr[x][]$.
We associate an approximate {\em cost} $\delta$ for accessing the store, with $c(\rd) \simeq c(\wr) \simeq \delta$.
For simplicity, we ascribe a uniform cost $\theta$ for $\tau$-labelled transitions (typically $\theta < \delta$ )\footnote{Assumptions on $\theta$ do not have any significance in this paper.}
  
In a sequence of transitions $\C_0 \trans{0} \cdots \trans{n} \C_{n+1}$, two {\em concurrently enabled but conflicting} transitions $\trans{j}$ and $\trans{k}$ are said to form a \emph{race} (on variable $x$) if $i_j \neq i_k$ and $a_j, a_k \in \{\rd[x][], \wr[x][]\}$ and at least one is $\wr[x][]$.
Races make computational results dependent on scheduling decisions, and as a consequence programmers use synchronisation mechanisms such as locks, barriers, fences, etc. to avoid the occurrence of such data races.

\section{Implementation Models}\label{CLUSTERINGS}

There is a variety of configurations for multicore architectures.  
Common among them are chip multiprocessors (CMP) and symmetric multiprocessing (SMP).  
A major difference between them lies in the organisation of their cache hierarchy.
Caches are  important architectural features that significantly speed up execution of programs by exploiting locality of memory accesses and reducing their latency.   

In a CMP configuration, each core possesses its private data and instruction $L1$ caches,  but several cores share a common $L2$ cache, which lies above the slower main memory.   
In contrast, in SMP, the $L2$ cache is also private to each core.
These two configurations may be considered the extremes of a range of {\em clustered configurations} or {\em clusterings}, where the multicore system consists of a collection of clustered cores.  
Within a cluster, each core possesses its private $L1$ data and instruction caches, but the cores share a $L2$ cache.  
Thus, SMP is the case where the cluster size is 1, whereas CMP puts all the cores in one cluster.   
Pal {\em et al.} use the notation $k(c_1, \ldots , c_k)$, where $(\Sigma_{j=1}^{k} c_j ) = N$ to describe a system of $N$ cores configured into $k$ clusters, where the $j^{th}$ cluster has $c_j$ cores.   
SMP is therefore written as $N(1, \ldots, 1)$ while CMP is represented as $1(N)$.   
For simplicity, we will only concern ourselves with a memory structure consisting of a $L2$ caches and shared main store.  A more detailed model can address the similar issues that arise in the treatment of $L1$ {\it vis \`{a} vis} $L2$ caches and main memory.

We observe that a clustering represents a {\em partition} of cores; given a clustering $Q$, we write $i \sim_Q j$ if cores $i$ and $j$ are in the same cluster.    
Thus in SMP, the equivalence classes are singletons, whereas in pure CMP, all cores are in the same equivalence class. 
If clustering $Q$ is a {\em partition refinement} of $Q'$, we write $Q \leq Q'$.\footnote{Note that refinements are lower in the ordering.}
Note that if $i \sim_Q j$, then $i,j$ share the same $L2$ cache. 

\subsection{Implementation semantics}\label{IMPLEMENTATION}

We now refine the reference model by introducing caches into the architecture.  
The store component  is replaced by a tuple $(S,C)$, where $S$ is the store (as earlier) and $C$ is a vector of caches.  
The caches contain a local copy of a subset of the store. 
Due to differences in the local caches, each core has a potentially different view of the memory.
$C_i$ denotes the ($L2$) cache available to core $i$.
In clustering $Q$, if $i \sim_Q j$, then $C_j$ and $C_i$ are the same and so have the same contents.

If $x \in \m{dom}(C_i)$,  its value $C_i[x]$ is given by a pair $(\m{val},\m{state})$, where $\m{val}$ is the value of the variable and $\m{state}$ may be either \clean{} or \dirty{}. 
A variable is \clean{} either if it has not been written to by \emph{this} cluster of cores, or if its changed value has been written through to the store. 
Otherwise it is \dirty{}. 
Note that in general, $C_i[x] = (v,\clean) \not\Rightarrow S(x) = v$.  The system may allow the store to contain a different value if some other processor has updated the store but this cache has not yet been notified. 
\begin{figure}[t]
\begin{tabularx}{\textwidth}{>{\(}l<{\)}}
(LocalRead)~~~(S, C, P_i[x])  \ptrans{\rd[x][v][l]}_Q  (S, C, P_i[v])  ~~~~~~~~~~~~~\textit{where } x \in \m{dom}(C_i) \land \val{i} = v\\
(StoreRead)~~~(S, C, P_i[x])  \ptrans{\rd[x][v][s]}_Q  (S, C, P_i[v]) ~~~~~~~~~~~~~\textit{where }x \not \in \m{dom}(C_i) \land S(x) = v\\
(ReadPull)~~~(S, C, P_i[x])  \ptrans{\rd[x][v][p]}_Q  (S, C_i[x \gets (v, \clean)], P_i[v])  ~~\textit{where }x \not \in \m{dom}(C_i) \land S(x) = v\\
(WriteBack)~~~(S, C, P_i[x $\assign$ v])  \ptrans{\wr_{[i]}}_Q (S, C_i[x \gets (v,\dirty)], P_i[()])  \\
\end{tabularx}
\caption{Implementation semantics for read and write operations on a clustering $Q$}\label{cachesem}
\end{figure}

Figure \ref{cachesem} gives the implementation semantics {\em with respect to clustering $Q$} for read and write operations, both of which access the memory --- and potentially alter it.
When a variable is written to, the write is only to the cache (``write back'').
We discuss below how these changes are propagated to the store or to other caches. 
Observationally, $\wr_{[i]}$ is a {\em functionally equivalent} action to $\wr$, but with lower cost: $c(\wr_{[i]}) \simeq \kappa < \delta$.

There are three transitions for reading a variable,  $\rd[x][v][l], \rd[x][v][s], \rd[x][v][p]$, all functionally equivalent to
the same specification operation $\rd$,  but with different costs.
$\rd[x][v][l]$ is a read from the local cache, and has cost $c(\rd[x][v][l])  \simeq \kappa$.
Note that when $x \not \in \m{dom}(C_i)$, there are two possible transitions, labelled
$\rd[x][v][s]$ and $\rd[x][v][p]$, both with costs $\delta$,
corresponding to whether or not $x$ is pulled into the cache. 
This decision is made non-deterministically, which (along with another transition for \emph{eviction} to be introduced later) makes the model independent of the \emph{cache-replacement} policy used by the actual implementation.
Note that unlike in the specification semantics, {\em reading} a location $x$ can cause changes to the memory, e.g., by moving the  value read into a cache. 

Apart from  the {\em programmed transitions} of Figures \ref{specificationsem} and \ref{cachesem},  there are the so-called  `system' transitions, denoted by $\xrightarrow{}_Q$, used to manage the memory structure, including cache replacement policies and consistency.  
These transitions can fire non-deterministically at any time, and the  threads cannot constrain which system transitions can occur or when. 
The system transitions are used to propagate writes to other caches and the store. 
In practice this is usually done either with an \emph{update-based} protocol (where cached copies are updated with the new value) or with an \emph{invalidation-based} protocol (where cached copies are invalidated, effectively removing them from the cache).  
Here we present only the update protocol (Figure \ref{updatesem}).   The transitions, which are not observable, but decorated here  to distinguish them, are as follows: 
\begin{figure}[t]
\begin{tabularx}{\textwidth}{>{\(}l<{\)}}
(Evict)~~~(S, C, P)  \eviction {}_Q ~ (S, C_i\uparrow x, P)  
~~~~~\hfill C_i[x] = (v, \clean) \\
(CacheUpd)~~~(S, C, P)  \cacheup{i}{j} {}_Q ~ (S, C_j[x \gets (v, \clean)], P)    \\ 
~~~~~~~~\hfill x \in \dom{j}\land C_j[x]\! \neq\! (v, \clean)\! \land\! C_i[x]\! =\! (v, \dirty) \\
(StoreUpd)~~~(S, C, P)  \storeup {}_Q ~ (S[x \gets v], C_i[x \gets (v,\clean)], P)  \\
~~~~~~~~~~~~~~~~~~~~~~ \hfill \forall j\! :\! j\! \neq\! i\! \land\! x\! \in\! \dom{j}, C_j[x]\! =\! (v, \clean)\! \land\! C_i[x]\! =\! (v, \dirty) \\
\end{tabularx}
\caption{System transitions on clustering $Q$: Update-based cache consistency protocol}\label{updatesem}
\end{figure}

\begin{enumerate}
\item \b{Eviction $\eviction$: } Evict $x$ from $C_i$. 
$c(\eviction)\simeq 0$. 
This is only used for the cache replacement policy and is not needed to achieve a consistent state. 
\item \b{Cache update $\cacheup{i}{j}$: } Update $x$ in $C_j$ from $C_i$.
$c(\cacheup{i}{j}) \simeq 0$ if  $i \sim_Q j$, since $C_i$ is the same as $C_j$; otherwise it is $\simeq \delta$ since it requires communication over the system bus.
This is used to update other caches when a variable is written to in a cache. 
\item \b{Store update $\storeup$: } Update $x$ in $S$ from $C_i$. 
$c(\storeup) \simeq \delta$.
The condition for its application ensures that a store update only happens \emph{after} all caches have been updated and agree on the value of the variable.
\end{enumerate}

\subsection{Comparing the semantics on different configurations}

Consider a program or workload  $W$, i.e., a set of threads mapped to a set of cores. 
An execution trace $\sigma$ of an implementation of $W$ on clustering $Q$ is considered correct if for any two actions $a,b \in \sigma$,  some functionally equivalent actions $a',b'$  both appear in the pomset semantics, and if $a'$ precedes $b'$ in the pomset semantics,  $a$ appears before $b$ in $\sigma$.  
The implementation conforms to the pomset semantics if every trace $\sigma$ possible in that implementation is correct with respect to the pomset semantics.
In the interleaving view, this can be stated as: for any observable trace in the reference semantics, there is a corresponding trace of functionally equivalent observable actions in the implementation semantics.   

Running any workload  $W$ on a coarser clustering preserves the observable behaviour.    
Moreover, CMP is semantically faithful to the reference semantics.
\begin{proposition}\label{PROP-COARSER}
	\begin{enumerate}
\item	If $Q' \leq Q$, then any $Q$-trace $\sigma$ has a functionally equivalent $Q'$-trace $\sigma'$.  
\item  Every reference semantics trace $\sigma$ has a functionally equivalent CMP-trace $\sigma'$ and vice versa.  
\end{enumerate}
\end{proposition}
\emph{Proof outline:} By induction on $\sigma$, we find a functionally equivalent $\sigma'$.
The only interesting cases are if $i \sim_Q j$ but $i \not\sim_{Q'} j$, and there is a local read to or write from (say)  $C_i[x]$.  We 
use the $\cacheup{i}{j} {}_{Q'}$ transition before a read or after a write to make the two cache entries agree.
Similarly, the $\storeup$ transition is used to make $C_i[x]$ agree with $S[x]$. \hfill $\square$

 \paragraph{Coherence, Consistency and Data Race Free programs.}\label{COHERENCE}
 
 Unfortunately, program execution on multiprocessors with caches may exhibit more traces than the reference model allows, due to the introduction of race conditions and inconsistencies between the caches at different cores or with the shared store (arising from the non-atomicity of writes).  
 It is therefore not true in general that program behaviour is  preserved when running a program on a finer clustering (e.g., SMP). 
 
 We recall below some of our earlier results showing that for a class of programs that are ``{\em data race free} (DRF)''  \cite{BoudolP-relaxed}, every program trace in any implementation architecture has a functionally equivalent trace in the reference architecture. 
 For such DRF programs, the additional behaviours, introduced by the extra nondeterminism in the implementation architecture, are irrelevant for executions starting from ``{\em consistent states}''.  
 A consistent state is, intuitively, an implementation state that is identifiable in a precise sense with a specific state (called its ``reduct'') in the reference model.
 
 We briefly recall the notions of coherence and consistency presented in our earlier work \cite{JoshiP-caches} via abstract operational characterisations. 
 We refer the interested reader to {\em op. cit.} to check that these notions correspond with more familiar invariants associated with memory consistency and cache coherence presented in the literature.  
 
 \noindent
 We write $\C \to^*_Q \C'$ to denote that $\C'$ is reachable from $\C$ by the \emph{implementation semantics} (program and system transitions) with respect to clustering $Q$, and similarly $\C \to^*_S \C''$ for the reference architecture semantics. We use $\strans {}_{\langle Q \rangle}$ to denote 0 or more \emph{system} transitions in clustering $Q$, whereas $\to^*_Q$ means 0 or more system and program transitions.
 
 \noindent
 Let us call a state $\C$ ``$\to_{\langle Q \rangle}$-normal'' if it cannot make any $\xrightarrow{}_{\langle Q \rangle}$ moves (i.e.\ system transitions).  For an implementation state $\C$, let $\C.M_i[x].\m{val}$ denote the value in core $i$'s  view of $x$,  i.e., the value of $C_i[x]$, or $S[x]$ if $x \not\in \dom{i}$.  An implementation state $\C$ is said to \emph{reduce to} a specification state $\C_S$ (written $\C \Downarrow_Q \C_S$) if $\exists \C': \C \strans {}_{\langle Q \rangle} \C'$, $\C'$ is $\to_{\langle Q \rangle}$-normal, and $\forall i \forall x\; \C'.M_i[x].\m{val} = \C_S.S[x]$. \; $\C_S$ is called a \emph{reduct} of $\C$.

 \begin{definition}
 	A state $\C$ is said to be \emph{coherent for $x$} if $\exists v : \forall i : x \in \dom{i} \land \state{i} = \dirty{} \Rightarrow \val{i} = v$.
 	A  state is \emph{coherent} if it is coherent for all $x$.   {\em A coherent state has a unique reduct}. We use $\reduct{\C}$ to refer to the unique reduct of a coherent state $\C$.
 \end{definition}
 
 \begin{definition}
 	A state $(S,C,P)$ is said to be \emph{consistent for $x$} if and only if $\forall i: x \in \dom{i}, C_i[x] = (S(x), \clean{})$.
 	Implementation state $\C$ is \emph{consistent} if it is consistent for all $x$.  A consistent state is in some sense identifiable with its reduct. 
 \end{definition}
 
 We now introduce the notion of data race freedom.
 
 \begin{definition}\label{drf-def}
 	A consistent state $\C$ involves a \emph{data race} if it has two redexes $P_i[r]$ and $P_j[r']$, $i \neq j$, $r$ and $r'$ are both accesses to the same variable and at least one is a write. $\C$ is \emph{data race free} (DRF) iff no state reachable in the reference architecture semantics from $\reduct{\C}$ involves a data race.
 \end{definition}

\noindent
Note also that the analysis of data race freedom need only be performed at the level of the reference  semantics. 
DRF programs allow us to consider their execution as progressing via a sequence of reference model states (reducts). 
Using  ideas and techniques introduced by Boudol and Petri \cite{BoudolP-relaxed}, it is shown that for ``well-synchronised'' programs, i.e., those where between any pair of actions forming a data race  lies an intervening synchronising mechanism (e.g., lock release or barrier operations), the behaviours are functionally equivalent. 
In particular, we have shown in \cite[Section 6]{JoshiP-caches} that the cache rules described above satisfy required properties of coherence and consistency.  

Since the execution of DRF programs on any clustering architecture can be seen to be functionally equivalent to execution on the reference architecture, we can show:
\begin{proposition}\label{PROP-DRF-CORRECT}
	For DRF programs, any reconfiguration (in any direction) preserves semantics.
\end{proposition}
\emph{Proof outline:}  Similar to the proof of Proposition \ref{PROP-COARSER}, but here we rely on the fact that every implementation trace of a DRF programs on clustering $Q$ has a functionally equivalent trace in the reference model.   The result follows from transitivity of equivalence of actions, and that `system' transitions are not directly observable. \hfill $\square$

\paragraph{Execution on a dynamically reconfiguring architecture.}

Consider now a scenario where execution may commence on clustering $Q$, and the machine may nondeterministically decide to morph to a clustering $Q'$ (based on some heuristic), after which execution proceeds on the latter architectural configuration.
Such {\em dynamic reconfiguration} from clustering $Q$ to $Q'$ can be 
internalised into our framework by introducing a new action $\reconf{Q}{Q'}$, with cost $c(\reconf{Q}{Q'}) = \mu \gg \delta$:
\[ (Reconf)~~~ \C ~\reconf{Q}{Q'} \ulcorner \C\urcorner
\]
where $\ulcorner \C\urcorner$ is the ``reduct'' of $\C$. 
On reconfiguration, the cache contents are written to the store, and the program resumes in the new configuration with ``cold caches''.
For DRF programs, the execution semantics of each phase corresponds precisely to the semantics of execution in the reference model.
Thus we can formalise within our framework the correctness criterion for execution on a reconfigurable architecture.  

\begin{theorem}\label{THM-RECONF-CORRECT}
	Any execution of a DRF program on a reconfigurable architecture conforms to execution on the reference model. 
\end{theorem}
\emph{Proof outline:}  By piecewise stitching of executions of the different phases on the different configurations.  Note that implementation states before and after reconfiguration correspond to the same reduct. \hfill $\square$

\section{Comparing Performance}

\noindent
We now propose a framework for comparing the execution efficiency on two configurations. 
For uniformity, specification and implementation states are clubbed into one set; 
the specification and implementation actions marking transitions are also combined into one set of actions ${\cal A}$.
Let $u \in {\cal A}^*$ be a sequence of actions, and let $\seqtrans{u}$ denote a labelled sequence of  transitions, as usual.

\noindent
Let $\rho \subset  {\cal A} \times {\cal A}$ represent the functional equivalence of actions as mentioned above.  
The actions $\rd[x][v][l], \rd[x][v][s], \rd[x][v][p], \rd$ are in one equivalence class; $\wr_{[i]}$ and $\wr$ in a second; and $\eviction, \cacheup{i}{j}, \storeup, \tau, \reconf{}{}$ in the third. 
For $\alpha$ in the read and write actions, let the observable content $\hat{\alpha} = \alpha$,
and for $\alpha \in \{\eviction, \cacheup{i}{j}, \storeup, \tau, \reconf{}{} \}$, define $\hat{\alpha} = \epsilon$ (the empty string).  
Extend $\rho$ to sequences such that $(\epsilon, \epsilon) \in \rho$ and $(\epsilon, u_1 ... u_n) \in \rho$ and $(u_1 ... u_n, \epsilon) \in \rho$ if
$(\tau, u_i) \in \rho$ for each $u_i$.

\noindent
Functionally equivalent actions deliver the same results but may have quite different latencies.  We have earlier specified the costs of actions $\kappa$ for cache accesses, $\delta$ for store accesses and $\mu$ for reconfiguration.    
Lift $c(~)$ to sequences by summing the costs of the component actions.  

\noindent
Following the constraints on $\rho$ and latency costs as in \cite{KiehnA-amortized}, we define the notion of weak amortised bisimulations on states (both specification and implementation)\footnote{A major difference with the cited work is that we ascribe a cost to observable actions as well}. 
\begin{definition}
	A family $(R_i)_{i \in \N}$ of
	binary relations over states is a {\em weak amortised $\rho$-bisimulation}, if for all $i \in \N$ whenever
	$(\C_1, \C_2) \in R_i$: \\
	$\C_1 \xrightarrow{a} \C_1'$ implies 
	$\exists \C_2', b, u, v: ~(a,b) \in \rho,  (\epsilon, u) \in \rho, (\epsilon, v) \in \rho, ~\C_2 \seqtrans{ u \hat{b} v } \C_2'$ and $(\C_1', \C_2') \in R_{i+c(u\hat{b}v)-c(a)}$, \\
	$\C_2 \xrightarrow{b} \C_2'$ implies 
	$\exists \C_1', a,  u, v: ~(a,b) \in \rho, (u, \epsilon) \in \rho,  (v, \epsilon) \in \rho, ~\C_1 \seqtrans{ u \hat{a} v }  \C_1'$ and $(\C_1', \C_2') \in R_{i+c(b)-c(u\hat{a}v)}$, \\
	where $a, b \in {\cal A}$ and $u, v \in {\cal A}^*$.   State $ \C_1$ is (weakly) amortised more efficient than $\C_2$ up to credit $i$,  written $\C_1 \preceq^{\rho}_i \C_2$, if $(\C_1, \C_2) \in R_i$ for some weak amortised
	$\rho$-bisimulation  $(R_i)_{i \in \N}$.  
\end{definition}
Note that accessing the cache is significantly less costly that accessing the store.  The definition of weak amortised bisimulation 
accumulates ``credit'' by performing the cheaper operation, thus providing us a framework for comparing the performance of execution on different architectural configurations.   Since there is nondeterminism in when the  `system' operations take place, our framework
must account for every possible execution run in the comparison.  

The notion also allows us to assess the benefits of performing dynamic reconfiguration.  
Reconfiguration introduces a handicap of $\mu$, which must be balanced (in an amortised sense) by frequent accesses to cache instead of to the store.  Typically $\delta$ is about 4 instruction cycles whereas $\kappa$ is one cycle.  Since values are pulled into cache in blocks of words, there are additional performance benefits due to overlapping reads with the execution of other operations.
The reconfiguration cost $\mu$ is approximately 1000 instruction cycles, so if there is enough locality of reference, this reconfiguration cost may be easily offset by the benefits of running the workload on a more suitable configuration for phases that are typically of the order of millions of instruction cycles or more.  

\begin{theorem}\label{THM-RECONF-EFF}
	\begin{enumerate}
		\item For DRF program states, any reconfiguration for more efficiency is permissible.  With sufficient locality of references, such programs executed on any clustering  are ``weakly amortised''  more efficient than execution under their reference model execution.
			\item If $Q \leq Q'$, then executing a program on configuration $Q'$ is (modulo the approximations on latency) ``weakly amortised more efficient'' than on $Q$.
	\end{enumerate}
\end{theorem}
\emph{Proof outline:} 
From any consistent  implementation state,  a unique reduct is reachable.   
If it takes $m \leq kn$  $\cacheup{i}{j}, \storeup$ moves to do so (where $k$ is the number of caches and $n$ the number of variables), then to have been in such a state, the system must have earlier performed  at least $m$ $\wr_{[i']}$ actions instead of $\wr$ actions, and thus have already ``earned'' a credit of $m (\delta - \kappa)$.   So if it has performed at least $m \kappa / (\delta - \kappa)$ operations such as distinct repeat writes to a variable in cache or reads of a $\dirty{}$ variable, then it has earned the requisite credit to be amortised at least as efficient as the reference system.  \\
For the second part, if $Q \leq Q'$, since there is more sharing of cached variables,  we can avoid the costlier $\rd[x][v][s], \rd[x][v][p]$  in favour of the cheaper $\rd[x][v][l]$ operation, and avoid some instances of the  $\cacheup{i}{j}$ operation. \hfill $\square$

\section{Conclusions}

Our framework provides a formal basis for comparing both the behaviour and the performance of program workloads on different multiprocessor architectures.  Our first set of results  (Propositions \ref{PROP-COARSER} and \ref{PROP-DRF-CORRECT})  provides us a rudimentary formal justification for the folklore that it is easy to port programs written assuming an SMP architecture to CMP.   In particular, they indicate why converting MPI programs to OpenMP, which assumes shared variables, is usually easier than the trickier reverse direction.  
Theorem \ref{THM-RECONF-CORRECT} indicates why as architectures become ``smarter'' and incorporate reconfigurability,  avoiding data races only assumes greater importance.  
The framework for comparing the efficiency of execution on different architecture also lets us understand why certain programs get such dramatic performance benefits when run on CMP architectures.  

Let us mention a shortcoming of our work.
Theorem \ref{THM-RECONF-EFF} seems to indicate that CMP is preferable to all other configurations, which is belied in reality, especially
where   applications with threads having little or no communication amongst themselves run more efficiently on SMP-like configurations.
The cache bus within a cluster and the memory bus enforce mutually exclusive access, and so the interleaved execution of threads accessing a shared cache is slower than  simultaneous independent execution of threads accessing private caches ($\kappa + \kappa$ interleaved vs $\kappa$ in parallel)\footnote{There also is a slowdown due to the fused cache being larger, which we neglect.}.
The fault lies not in our framework, but rather in our view of execution as being interleaved --- a view we had taken to keep the semantics standard.  We leave for the future the development of a framework where the pomset model is used to explore the correctness and efficiency issues.  

Reconfiguration is also an opportunity for remapping threads to cores.  
While we have considered only a fixed workload mapping of threads to cores in the present paper,  we do not believe this extension poses any major technical difficulties.

\paragraph{Acknowledgements.} I wish to acknowledge the helpful discussions with my colleague Kolin Paul who taught me the little I know about reconfigurable architectures.  I also must thank one of the referees who constructively pointed out many major weaknesses of the earlier avatar of this paper.  

\bibliographystyle{eptcs}
\bibliography{multi-bib}
\end{document}